\newcommand\kms{km s$^{-1}$}
\newcommand\aj{{AJ\,}}%
\newcommand\araa{{ARA\&A\,}}%
\newcommand\apj{{ApJ\,}}%
\newcommand\apjl{{ApJ\,}}%
\newcommand\apjs{{ApJS\,}}%
\newcommand\aap{{A\&A\,}}%
\newcommand\mnras{{MNRAS\,}}%
\newcommand\pasa{{PASA\,}}%
\documentclass[graybox, natbib, footinfo]{svmult}
\usepackage{mathptmx}       % selects Times Roman as basic font
\usepackage{helvet}         % selects Helvetica as sans-serif font
\usepackage{courier}        % selects Courier as typewriter font
\usepackage{type1cm}        % activate if the above 3 fonts are
\usepackage{makeidx}         % allows index generation
\usepackage{graphicx}        % standard LaTeX graphics tool
\usepackage{multicol}        % used for the two-column index
\usepackage[bottom]{footmisc}% places footnotes at page bottom
\usepackage{url}

\makeindex             % used for the subject index
\begin{document}

\title*{Structure and Evolution of the Milky Way}
\author{Ken Freeman}
\institute{K. Freeman \at Australian National University, Canberra, Australia \email{kcf@mso.anu.edu.au}}
\maketitle

\abstract{This review discusses the structure and evolution of the
Milky Way, in the context of opportunities provided by asteroseismology
of red giants. The review is structured according to the main Galactic
components: the thin disk, thick disk, stellar halo, and the Galactic
bar/bulge.  The review concludes with an overview of Galactic
archaeology and chemical tagging, and a brief account of the upcoming
HERMES survey with the AAT.}

\section{The Thin Disk: Formation and Evolution}
Here are some of the issues related to the formation and evolution of 
the Galactic thin disk:
\begin{itemize}
\item Building the thin disk: its exponential radial structure, and the 
role of mergers.
\item The star formation history: chemical evolution and continued gas accretion.
\item Evolutionary processes in the disk: disk heating, radial mixing.
\item The outer disk: chemical properties and chemical gradients.
\end{itemize}

Many of the basic observational constraints on the properties of
the Galactic disk are still uncertain. At this time, we do not have
reliable information about the star formation history of the disk.
We do not know how the metallicity distribution and the stellar
velocity dispersions in the disk have evolved with time. One might
have expected that these observational questions were well understood
by now, but this is not yet so.  The basic observational problem
is the difficulty of measuring ages for individual stars.

The younger stars of the Galactic disk show a clear abundance
gradient of about 0.07 dex kpc$^{-1}$, outlined nicely by the
cepheids \citep{Luck2006}.  In the outer disk, for the older
stars, the abundance gradient appears to be even stronger: the
abundance gradient (and the gradient in the ratio of alpha-elements
to Fe) have flattened with time towards the solar values.  A striking
feature of the radial abundance gradient in the Galaxy is that it
flattens for $R > 12$ kpc at an [Fe/H] value of about -0.5 \citep{Carney2005}.
  A similar flattening of the abundance gradient is
seen in the outer regions of the disk of M31 \citep{Worthey2005}.

The relation between the stellar age and the mean metallicity and
velocity dispersion are the fundamental observables that constrain
the chemical and dynamical evolution of the Galactic thin disk.
The age-metallicity relation (AMR) in the solar neighborhood is still
uncertain. Different authors find different relations, ranging from
a relatively steep decrease of metallicity with age  from \cite{Rocha-Pinto2004}
 to almost no change of mean metallicity with age from
\cite{Nordstrom2004}.  Much of the earlier work indicated that
a large scatter in metallicity was seen at all ages, which was
part of the motivation to invoke large-scale radial mixing of stars
within the disk. This mixing, predicted by \cite{Sellwood2002}, 
is generated by resonances with the spiral pattern, and is
able to move stars from one near-circular orbit to another.  It
would bring stars from the inner and outer disks, with their different
mean abundances, into the solar neighborhood.  Radial mixing is
potentially an important feature of the evolution of the disk.  At
this stage, it is a theoretical concept, and it is not known how
important it is in the Galactic disk. We are not aware of any strong
observational evidence at this stage for the existence of radial
mixing.  More recent results on the AMR (e.g.  Wylie de Boer \etal,
unpublished) indicate that there is a weak decrease of mean metallicity
with age in the Galactic thin disk, but that the spread in metallicity
at any age is no more than about 0.10 dex. If this is correct, then
radial mixing may not be so important for chemically mixing the
Galactic disk.

The age-velocity dispersion relation (AVR) is also not well determined
observationally. The velocity dispersion of stars appears to increase
with age, and this is believed to be due to the interaction of stars
with perturbers such as giant molecular clouds and transient spiral
structure. But there is a difference of opinion about the duration
of this heating.  One view is that the stellar velocity dispersion
$\sigma$ increases steadily for all time,  $\sim t^{0.2-0.5}$, based
on \cite{Wielen1977}'s work using chromospheric ages and kinematics for
the McCormick dwarfs.  Another view \citep[e.g.][]{Quillen2000},
based on the data for subgiants from \cite{Edvardsson1993} is that
the heating takes place for the first $\sim 2$ Gyr, but then saturates
when $\sigma \approx 20$ km s$^{-1}$ because the stars of higher
velocity dispersion spend most of their orbital time away from the
Galactic plane where the sources of heating lie. Data from \cite{Soubiran2008}
 support this view.  Again, much of the difference in view
goes back to the difficulty of measuring stellar ages.  Accurate ages
from asteroseismology would be very welcome.  Accurate ages and distances 
for a significant sample of red giants would allow us to measure the
AMR and AVR out to several kpc from the Sun.  This would be a great step
forward in understanding the chemical and dynamical evolution of the
Galactic disk.

\section{The Formation of the Thick Disk}
Most spiral galaxies, including out Galaxy, have a second thicker disk
component. For example, the thick disk and halo of the edge-on spiral galaxy
NGC 891, which is much like the Milky Way in size and morphology, has a 
thick disk nicely seen in star counts from HST images \citep{Mouhcine2010}. 
Its thick disk has scale height $\sim 1.4$ kpc and scalelength $\sim 4.8$ kpc, 
much as in our Galaxy. The fraction of baryons in the thick disk is typically
about $10$ to $15$ percent in large systems like the Milky Way, but rises to 
about $50$\% in the smaller disk systems \citep{Yoachim2008}. 

The Milky Way has a significant thick disk, discovered by \cite{Gilmore1983}.
 Its vertical velocity dispersion is about 40 \kms;
its scale height is still uncertain but is probably about $1000$ pc.  The
surface brightness of the thick disk is about 10\% of the thin
disk's, and near the Galactic plane it rotates almost as rapidly
as the thin disk.  Its stars are older than 10 Gyr and are significantly
more metal poor than the stars of the thin disk; most of the thick
disk stars have [Fe/H] values between about $-0.5$ and $-1.0$ and
are enhanced in alpha-elements relative to Fe.  This is usually interpreted
as evidence that the thick disk formed rapidly, on a timescale $\sim 1$ Gyr.
From its kinematics and chemical properties, the thick disk appears to be
a discrete component, distinct from the thin disk. Current opinion is that the
thick disk shows no vertical abundance gradient \citep[e.g.][]{Gilmore1995,
Ivezic2008}.  

The old thick disk is a very significant component for studying
Galaxy formation,  because it presents a kinematically and chemically
recognizable relic of the early Galaxy.  Secular heating is unlikely
to affect its dynamics significantly, because its stars spend most
of their time away from the Galactic plane.

How do thick disks form ? Several mechanisms have been proposed, including:
\begin{itemize}
\item thick disks are a normal part of early disk settling, and
form through energetic early star forming events, e.g. in gas-rich
mergers \citep{Samland2003, Brook2004}
\item thick disks are made up of accretion debris \citep{Abadi2003}.
  From the mass-metallicity relation for galaxies, the accreted 
galaxies that built up the thick disk of the Galaxy would need to be more 
massive than the SMC to get the right mean [Fe/H] abundance ($\sim -0.7$). 
The possible discovery of a counter-rotating thick disk \citep{Yoachim2008}
 in an edge-on galaxy would favor this mechanism.
\item thick disks come from the heating of the thin disk via
disruption of its early massive clusters \citep{Kroupa2002}. The internal
energy of large star clusters is enough to thicken the disk. 
Recent work on the significance of the high redshift clump structures
may be relevant to the thick disk problem: the thick disk may
originate from the merging of clumps and heating by clumps \citep[e.g.][]{Bournaud2009}.
 These clumps are believed to form by gravitational
instability from turbulent early disks: they appear to generate thick disks
with scale heights that are radially approximately uniform, rather than 
the flared thick disks predicted from minor mergers.
\item thick disks come from early partly-formed thin disks, heated 
by accretion events such as the accretion event which is believed to have
brought omega Centauri into the Galaxy \citep{Bekki2003}. In this
picture, thin disk formation began early, at $z = 2$ to $3$.  The
partly formed thin disk is partly disrupted during the active merger epoch 
which heats it into thick disk observed now, The rest of the gas then 
gradually settles to form the present thin disk, a process which continues
to the present day.
\item  a recent suggestion is that stars on more energetic orbits migrate 
out from the inner galaxy to form a thick disk at larger radii where the 
potential gradient is weaker \citep{Schonrich2009} 
\end{itemize}

How can we test between these possibilities for thick disk formation?  
\cite{Sales2009} looked at the expected orbital eccentricity
distribution for thick disk stars in different formation scenarios.
Their four scenarios are:
\begin{itemize}
\item a gas-rich merger: the thick disk stars are born in-situ
\item  the thick disk stars come in from outside via accretion
\item the early thin disk is heated by accretion of a massive satellite
\item the thick disk is formed as stars from the inner disk migrate out
to larger radii.
\end{itemize}
Preliminary results from the observed orbital eccentricity distribution for
thick disk stars may favor the
gas-rich merger picture \citep{Wilson2011}. This is
a potentially powerful approach for testing ideas about the origin of the
thick disk.  Because it depends on the orbital properties of the thick
disk sample, firm control of selection effects is needed in the 
identification of which stars belong to the thick disk. Kinematical
criteria for choosing the thick disk sample are clearly not ideal.

To summarize this section on the thick disk:  Thick disks are very common
in disk galaxies.  In our Galaxy, the thick disk is old, and is
kinematically and chemically distinct from the thin disk.  It is
important now to identify what the thick disk represents in the galaxy
formation process.  The orbital eccentricity distribution of the thick
disk stars will provide some guidance.  Chemical tagging will show if 
the thick disk formed as a small number of very large aggregates, or 
if it has a significant contribution from accreted galaxies.  This is 
one of the goals for the upcoming AAT/HERMES survey: see section 5.

\section{The Galactic Stellar Halo} The stars of the Galactic halo
have [Fe/H] abundances mostly less than -1.0.  Their kinematics are
very different from the rotating thick and thin disks: the mean
rotation of the stellar halo is close to zero, and it is supported
against gravity primarily by its velocity dispersion.  It is now
widely believed that much of the stellar halo comes from the debris
of small accreted satellites \citep{Searle1978}. There remains
a possibility that a component of the halo formed dissipationally
during the Galaxy formation process \citep{Eggen1962, Samland2003}.
  Halo-building accretion events continue to the
present time: the disrupting Sgr dwarf is an example in our Galaxy,
and the faint disrupting system around NGC 5907 is another example
of such an event \citep{Martinez-Delgado2010}.  The metallicity distribution
function (MDF) of the major surviving satellites around the Milky
way is not like the MDF in the stellar halo \citep[e.g.][]{Venn2008}
but the satellite MDFs may have been more similar long ago.  We
note that the fainter satellites are more metal-poor and are
consistent with the Milky Way halo in their [$\alpha$/Fe] behaviour.

Is there a halo component that formed dissipationally early in the
Galactic formation process?  \cite{Hartwick1987} showed that the
metal-poor RR Lyrae stars delineate a two-component halo, with a
flattened inner component and a spherical outer component.
\cite{Carollo2010} identified a two-component halo and the thick
disk in a sample of 17,000 SDSS stars, mostly with [Fe/H] $< -0.5$.
They described the kinematics well with these three components:\\
\hspace*{5mm}Thick disk: ($\bar{V}, \sigma$, [Fe/H]) = (182, 51, -0.7)\\
\hspace*{5mm}Inner halo: ($\bar{V}, \sigma$, [Fe/H]) = (7, 95, -1.6)\\
\hspace*{5mm}Outer halo: ($\bar{V}, \sigma$, [Fe/H]) = (-80, 180, -2.2) \\
Here [Fe/H] is the mean abundance for the component, $\bar{V}$ and $\sigma$ 
are its mean rotation velocity relative to
a non-rotating frame, and velocity dispersion, in \kms. The outer
halo appears to have retrograde mean rotation. 
As we look at subsamples at greater distances from the Galactic plane,
we see that the thick disk dies away and the retrograde outer halo takes over
from the inner halo. With the above kinematic parameters, the equilibrium 
of the inner halo is a bit hard to understand. It may not yet be in 
equilibrium.  From comparison with simulations, \cite{Zolotov2009} 
argue that the inner halo has a partly dissipational origin, while the
outer halo is made up from debris of faint metal-poor accreted satellites.

Recently \cite{Nissen2010} studied a sample of 78 halo stars
with [Fe/H] $>-1.6$ and find that they show a variety of [$\alpha$/Fe]
enhancement. Their sample shows high and low [$\alpha$/Fe] groups,
and the low [$\alpha$/Fe] stars are mostly in high energy retrograde
orbits.  The high [$\alpha$/Fe] stars could be ancient halo stars
born in situ and possible heated by satellite encounters.  The
low-alpha stars may be accreted from dwarf galaxies.

How much of the halo comes from accreted structures?  An ACS study
by \cite{Ibata2009} of the halo of NGC 891 (a nearby edge-on
galaxy like the Milky Way) shows a spatially lumpy metallicity
distribution, indicating that its halo is made up largely of accreted
structures which have not yet mixed away.  This is consistent with
simulations of stellar halos by \cite{Font2008}, \cite{Gilbert2009}
 and \cite{Cooper2010}.

To summarize this section on the Galactic stellar halo: the stellar
halo is probably made up mainly of the debris of small accreted
galaxies, although there may be an inner component which formed
dissipatively.

\section{The Galactic Bar/Bulge}
The boxy appearance of the Galactic bulge is typical of galactic bars 
seen edge-on. These bar/bulges are very common: about 2/3 of spiral 
galaxies show some kind of central bar structure in the infra-red. 
Where do these bar/bulges come from ?

Bars can arise naturally from the instabilities of the disk. A
rotating disk is often unstable to forming a flat bar structure at
its center.  This flat bar in turn is often unstable to vertical
buckling which generates the boxy appearance. This kind of bar/bulge 
is not generated by mergers but follows simply from the dynamics of
a flat rotating disk of stars. The maximum vertical extent of boxy or
peanut-shaped bulges occurs near the radius of the vertical and 
horizontal Lindblad resonances, i.e. where 
$$\Omega_b = \Omega - \kappa/2 = \Omega - \nu_z/2.$$  
Here $\Omega$ is the circular angular velocity, $\Omega_b$ is the
pattern speed of the bar, $\kappa$ is the epicyclic frequency and
$\nu_z$ is the vertical frequency of oscillation. We note that the frequencies 
$\kappa$ and particularly $\nu_z$ depend on the amplitude of the
oscillation. Stars in this zone oscillate on 3D orbits which support
the peanut shape.

We can test whether the Galactic bulge formed through this kind of
bar-buckling instability of the inner disk, by comparing the structure
and kinematics of the bulge with those of N-body simulations that 
generate a boxy/bar bulge \citep[e.g.][]{Athanassoula2005}. The simulations show
an exponential structure and near-cylindrical rotation: do these
simulations match the properties of the Galactic bar/bulge?

The stars of the Galactic bulge appear to be old and enhanced in
$\alpha$-elements. This implies a rapid history of star formation.
If the bar formed from the inner disk, then it would be interesting
to know whether the bulge stars and the stars of the adjacent disk
have similar chemical properties.  This is not yet clear.  There
do appear to be similarities in the $\alpha$-element properties
between the bulge and the thick disk in the solar neighborhood \citep[e.g.][]
{Melendez2008}.

The bar-forming and bar-buckling process takes 2-3 Gyr to act after
the disk settles.  In the bar-buckling instability scenario, the
bulge {\it structure} is probably younger than the bulge {\it stars},
which were originally part of the inner disk.  The alpha-enrichment
of the bulge and thick disk comes from the rapid chemical evolution
which took place in the inner disk before the instability acted.
In this scenario, the stars of the bulge and adjacent disk should 
have similar ages: accurate asteroseismology ages for giants of the
bulge and inner disk would be a very useful test of the scenario.

We are doing a survey of about 28,000 clump giants in the Galactic
bulge and the adjacent disk, to measure the chemical properties 
(Fe, Mg, Ca, Ti, Al, O) of stars in the bulge and adjacent disk: are 
they similar, as we would expect if the bar/bulge grew out of the disk?   
We use the AAOmega fiber spectrometer on the AAT, to acquire 
medium-resolution spectra of about 350 stars at a time, at a
resolution $R \sim 12,000$.

The central regions of our Galaxy are not only the location of the
bulge and inner disk, but also include the central regions of the
Galactic stellar halo.  Recent simulations \citep[e.g.][]{Diemand2005,
Moore2006, Brook2007} indicate that the {\it metal-free}
(population III) stars formed until redshift $z \sim 4$, in chemically
isolated subsystems far away from the largest progenitor.  If its stars
survive, they are spread throughout the Galactic halo. If they are not
found, then it would be likely that their lifetimes are less than a
Hubble time which in turn implies a truncated IMF. On the other hand,
the {\it oldest} stars form in the early rare high density peaks that
lie near the highest density peak of the final system. They are not
necessarily the most metal-poor stars in the Galaxy. Now, these oldest
stars are predicted to lie in the central bulge region of the Galaxy. 
Accurate asteroseismology ages for metal-poor stars in the inner Galaxy  
would provide a great way to tell if they are the oldest stars or just 
stars of the inner Galactic halo. This test would require a $\sim 10$\%
precision in age.

Our data so far indicate that the rotation of the Galactic bulge
is close to cylindrical \citep[see also][]{Howard2009}.  Detailed
analysis will be needed to see if there is any evidence for a small
classical merger generated bulge component, in addition to the
boxy/peanut bar/bulge which probably formed from the disk.  We also
see a more slowly rotating metal-poor component in the bulge region.
The problem now is to identify the {\it first} stars from among the
expected metal-poor stars of the inner halo.

\section{Galactic Archaeology}
The goals of Galactic Archaeology are to find signatures or fossils from
the epoch of Galaxy assembly, to give us insight into the processes that
took place as the Galaxy formed.  A major goal is to identify observationally
how important mergers and accretion events were in building up the Galactic 
disk, bulge and halo of the Milky Way.  CDM simulations predict a high
level of merger activity which conflicts with some observed properties
of disk galaxies, particularly with the relatively common nature of
large galaxies like ours with small bulges \citep[e.g.][]{Kormendy2010}.

The aim is to reconstruct the star-forming aggregates and accreted
galaxies that built up the disk, bulge, and halo of the Galaxy.
Some of these dispersed aggregates can still be recognized kinematically
as stellar moving groups.  For others, the dynamical information
was lost through heating and mixing processes, but their debris can
still be recognized by their chemical signatures (chemical tagging).
We would like to find groups of stars, now dispersed, that were
associated at birth either \begin{itemize} \item because they were
born together and therefore have almost identical chemical abundances
over all elements \citep[e.g.][]{deSilva2009}, or \item because
they came from a common accreted galaxy and have abundance patterns
that are clearly distinguished from those of the Galactic disk \citep[e.g.][]
{Venn2008}. \end{itemize}

The galactic disk shows kinematical substructure in the solar
neighborhood: groups of stars moving together, usually called moving
stellar groups.  Some are associated with dynamical resonances (e.g.
the Hercules group): in such groups, we do not expect to see chemical
homogeneity or age homogeneity \citep[e.g.][]{Bensby2007}. Others
are the debris of star-forming aggregates in the disk  (e.g. the
HR1614 group and Wolf 630 group).  They are chemically homogeneous,
and such groups could be useful for reconstructing the history of
the galactic disk.  Yet others may be debris of infalling objects,
as seen in CDM simulations \citep[e.g.][]{Abadi2003}.

The stars of the HR 1614 group appear to be the relic of a dispersed
star-forming event. These stars have an age of about 2 Gyr and
[Fe/H] $= +0.2$, and they are scattered all around us.  This group
has not lost its dynamical identity despite its age. \cite{deSilva2007}
 measured accurate differential abundances for many
elements in HR 1614 stars, and found a very small spread in abundances.
This is encouraging for recovering dispersed star forming events
by chemical tagging.

Chemical studies of the old disk stars in the Galaxy can help
to identify  disk stars which came in from outside in disrupting
satellites, and also those that are the debris of dispersed
star-forming aggregates like the HR 1614 group  \citep{Freeman2002}.
  The chemical properties of surviving satellites
(the dwarf spheroidal galaxies) vary from satellite to satellite,
but are different in detail from the overall chemical properties of 
the disk stars.

We can think of a chemical space of abundances of elements: O, Na,
Mg, Al, Ca, Mn, Fe, Cu, Sr, Ba, Eu for example.  Not all of these
elements vary independently. The dimensionality of this space
chemical space is probably between about 7 and 9.   Most disk stars
inhabit a sub-region of this space.  Stars that come from dispersed
star clusters represent a very small volume in this space. Stars
which came in from satellites may have a distribution in this space
that is different enough to stand out from the rest of the disk
stars.  With this chemical tagging approach, we hope to detect or
put observational limits on the satellite accretion history of the
galactic disk.

Chemical studies of the old disk stars in the Galaxy can identify 
disk stars that are the debris of common dispersed star-forming 
aggregates.  Chemical tagging will work if \begin{itemize}
\item stars form in large aggregates, which is believed to be true
\item aggregates are chemically homogenous
\item aggregates have unique chemical signatures defined by several 
elements or element groups which do not vary in lockstep from one 
aggregate to another.   We need sufficient spread in abundances 
from aggregate to aggregate so that chemical signatures can be 
distinguished with accuracy achievable ($\sim 0.05$ dex differentially)
\end{itemize}

de Silva's work on open clusters was aimed at testing the last two 
conditions: they appear to be true.  See \cite{deSilva2009} for 
more on chemical tagging.

We should stress here that chemical tagging is not just assigning
stars chemically to a particular population, like the thin disk,
thick disk or halo.  Chemical tagging is intended to assign stars
chemically to substructure which is no longer detectable kinematically.
We are planning a large chemical tagging survey of about a million
stars, using the new HERMES multi-object spectrometer on the AAT.
The goal is to reconstruct the dispersed star-forming aggregates that 
built up the disk, thick disk and halo within about 5 kpc of the sun.

HERMES is a new high resolution multi-object spectrometer on the AAT.
Its spectral resolution is about 28,000, with a high resolution mode
with $R = 50,000$. It is fed by 400 fibers over a 2-degree field, and
has 4 non-contiguous wavelength bands covering a total of about 1000\AA.
The four wavelength bands were chosen to include measurable lines of
elements needed for chemical tagging.  HERMES is scheduled for first
light in late 2012.  The HERMES chemical tagging survey will include 
stars brighter than $V = 14$ and has a strong synergy with Gaia: for
the dwarf stars in the HERMES sample, the accurate ($1$\%) parallaxes
and proper motions will be invaluable for more detailed studies.

The fractional contribution of the different Galactic components to 
the HERMES sample will be about $78$\% thin disk stars, $17$\% thick
disk stars and about $5$\% halo stars. About $70$\% of the stars
will be dwarfs within about 1000 pc and $30$\% giants within about 5 kpc.
About $9$\% of the thick disk stars and about $14$\% of the thin disk
stars pass within our 1 kpc dwarf horizon. Assume that all of their
formation aggregates are now azimuthally mixed right around the Galaxy,
so that all of their formation sites are represented within our horizon.
Simulations \citep{Bland-Hawthorn2004} show that a complete
random sample of about a million stars with $V < 14$ would allow
detection of about 20 thick disk dwarfs from each of about 4500 star
formation sites, and about 10 thin disk dwarfs from each of about
35,000 star formation sites.  These estimates depend on the adopted
mass spectrum of the formation sites. In combination with Gaia, HERMES
will give the distribution of stars in the multi-dimensional\{position, 
velocity, chemical\} space, and isochrone ages for about 200,000 stars
with $V < 14$.  We would be interested to explore further what the
HERMES survey can contribute to asteroseismology.

Some authors have argued that the thick disk may have formed from
the debris of the huge and short-lived star formation clumps observed
in disk galaxies at high redshift \citep[e.g.][]{Bournaud2009, Genzel2011}.
 If this is correct, then only a small number of these
huge building blocks would have been involved in the assembly of
the thick disk, and their debris should be very easy to identify via
chemical tagging techniques.

Chemical tagging in the inner regions of the Galactic disk will be
of particular interest. We expect about 200,000 survey giants in
the inner region of the Galaxy.  The surviving old ($> 1$ Gyr) open
clusters are all in the outer Galaxy, beyond a radius of 8 kpc.
Young open clusters are seen in the inner Galaxy, but do not appear
to survive the disruptive effects of the tidal field and giant
molecular clouds in the inner regions.  We expect to find the debris
of many broken open and globular clusters in the inner disk. These
will be good for chemical tagging recovery using the HERMES giants.
The radial extent of the dispersal of individual broken clusters will provide 
an acute test of radial mixing theory within the disk. Another 
opportunity comes from the the Na/O anomaly, which is unique to
globular clusters, and may help to identify the debris of disrupted
globular clusters.

\end{document}